\documentclass[twocolumn,showpacs,superscriptaddress,preprintnumbers,amsmath,amssymb,prc]{revtex4-1}
\usepackage{multirow}
\usepackage{amssymb}
\usepackage{amsmath}
\usepackage{graphicx}
\usepackage[normalem]{ulem}
\usepackage{cancel}
\usepackage{multirow}
\usepackage{CJK}
\usepackage[usenames]{color}
\usepackage[colorlinks,linkcolor=blue,urlcolor=blue,citecolor=blue]{hyperref}

\usepackage{tikz,xcolor,hyperref}

\definecolor{lime}{HTML}{A6CE39}
\DeclareRobustCommand{\orcidicon}{
	\begin{tikzpicture}
	\draw[lime, fill=lime] (0,0) 
	circle [radius=0.16] 
	node[white] {{\fontfamily{qag}\selectfont \tiny ID}};
	\draw[white, fill=white] (-0.0625,0.095) 
	circle [radius=0.007];
	\end{tikzpicture}
	\hspace{-2mm}
}
\foreach \x in {A, ..., Z}{%
	\expandafter\xdef\csname orcid\x\endcsname{\noexpand\href{https://orcid.org/\csname orcidauthor\x\endcsname}{\noexpand\orcidicon}}
}

\usepackage{soul,xcolor}
\setstcolor{red}
\newcommand\redsout{\bgroup\markoverwith{\textcolor{red}{\rule[0.5ex]{2pt}{0.4pt}}}\ULon}

\begin{document}
\begin{CJK*} {UTF8} {gbsn}

\title{``Soft" interaction parameters setting in the extended quantum
   molecular dynamics model}

\author{Chen-Zhong Shi(施晨钟)}
\affiliation{Shanghai Institute of Applied Physics,  Chinese Academy of Sciences, Shanghai 201800, China}
 \affiliation{Key Laboratory of Nuclear Physics and Ion-Beam Application (MOE), Institute of Modern Physics, Fudan University, Shanghai 200433, China}
\author{Xiang-Zhou Cai(蔡翔舟)}%
\affiliation{Shanghai Institute of Applied Physics,  Chinese Academy of Sciences, Shanghai 201800, China}%

\author{Bo-Song Huang(黄勃松)}%
\affiliation{Shanghai Institute of Applied Physics,  Chinese Academy of Sciences, Shanghai 201800, China}%

\author{Yu-Gang Ma(马余刚)\orcidC{}} \thanks{Corresponding author:  mayugang@fudan.edu.cn}
\affiliation{Key Laboratory of Nuclear Physics and Ion-Beam Application (MOE), Institute of Modern Physics, Fudan University, Shanghai 200433, China}
\affiliation{Shanghai Research Center for Theoretical Nuclear Physics， NSFC and Fudan University, Shanghai 200438, China }


\date{\today}             

\begin{abstract}
The extended quantum molecular dynamics (EQMD) model is one of the few quantum molecular dynamics (QMD)-like transport approaches that can be used to study the effective clustering structure as well as heavily deformed nuclei in both ground state nuclei and nuclear reactions. 
However, there are only two parameter sets that lead to hard incompressibility for long times.
The aim of the present work is to obtain a soft equation of state (EoS) in the EQMD model. 
In this context, we take the isoscalar giant monopole resonance (ISGMR), which is sensitive to the EoS, as an example to check our work. 
By introducing a kind of standard Skyrme energy density functional with different parameter sets, such as SkP, SkT1, and SKXce, whose incompressibility value ranges from 200 to 268 MeV, the ISGMR of $^{208}$Pb and other nuclei are studied.
When the SkP parameter sets are adopted, our new soft interaction in the EQMD model gives reasonable agreement with the experimental data in the heavy ion regime.
\end{abstract}

\maketitle

\section{Introduction\label{introduction}}
The low- to medium-energy heavy-ion reaction has served as a well-known tool to probe the structures of the atomic nucleus. Recently, the interest has been extended to much higher energies, the potential for probing nuclear deformation \cite{Ma_book,zhangchunjian,jiajiangyong} and cluster configurations \cite{ZhangS,liyian,hejunjie,MaL,CaoRX,WangYZ,Ma_NT,Ma_SCP} has been widely discussed, and several experimental evidences of nuclear deformation have been observed \cite{evidences1,evidences2,evidences3}.
In terms of transport models, the extended quantum molecular dynamics (EQMD) model \cite{EQMD} is one of the models used to study the nuclear cluster structures \cite{hewb1,hewb2,He1,He2} and the heavily deformed nucleus \cite{wangss_deformed}. 
Compared with most quantum molecular dynamics (QMD)-like models 
\cite{Aichelin,Buss,Wolter,MaCW}, 
the EQMD model developed by Maruyama {\it et al.} has been improved in the following aspects \cite{EQMD}. 
A phenomenological Pauli potential is adopted, a frictional cooling method is used to initialize the nucleus, the dynamical treatment of the wave packet width and the subtraction of the spurious zero-point kinetic energy of the fragment center of mass have been taken into account.
These improvements allow the EQMD model to study heavy ion collisions (HICs) near the Fermi energy region.   
In particular, for the currently concerned $\alpha$-clustering structure in light nuclei \cite{RMP, Nature, hewb2, ZhouB_NC} and deformed nucleus \cite{wangss_deformed}, EQMD can provide an opportunity to study the internal structure, the effects of $\alpha$-clustering and the deformation parameters in the nucleus-nucleus collision.
With reasonable computational performance, the EQMD model becomes one of the important transport models to study the exotic effects in low and medium energy nuclear reactions \cite{Ma_NT, Ma_book}.
Recently, the EQMD model has been used to study collective flow \cite{guocc}, giant dipole resonance \cite{hewb1,hewb2}, bremsstrahlung \cite{shicz1,shicz2}, photon-nuclear reaction \cite{huangbs}, shear viscosity \cite{guocq,wangk,DengXG}, short-range correlation \cite{shenl}, and the electromagnetic field effect \cite{caoyt} in nuclear reactions. 
However, in the following context it can be seen that both sets of effective interaction parameters used in EQMD~\cite{EQMD} only lead to hard infinite nuclear incompressibility coefficients ($K_{\infty}$).
This confuses with a current consensus, i.e. the $K_\infty$ is in the range of $230 \pm 40$ MeV \cite{khan}.
The incompressibility coefficient is a key parameter in the nuclear equation of state~\cite{youngblood, Tin1, Tin2, umesh, NST_An, PRC_Khan, SCPMA_LI, WangR, NST_Xie, NST_Xu, Colo23} and in the astrophysics of the core-collapse supernova and the neutron star.
It could also affect the treatment of nuclear reactions such as bremsstrahlung \cite{SCHUTZ1997404}, collective flow \cite{Ogilvie,flow_eos}, the positively charged kaon yields \cite{kaon}, and so on.  
This is crucial to the accuracy of our studies of exotic structure in low to medium energy nuclear reactions. In this context 
the situation where $K_{\infty}$ is too large in the original EQMD model is worth addressing.

As a sensitive observable for extracting valuable information about the incompressibility coefficient, the isoscalar giant monopole resonance (ISGMR), known as the breathing mode, has been extensively studied in the last few decades \cite{first,youngblood,Youngblood2,Youngblood3,Vand,umesh,Colo}.
For example, $^{208}$Pb as a double closed shell nucleus, its ISGMR behavior has been studied in detail to extract the nuclear incompressibility value \cite{umesh,Piekarewicz2}.
The series of Sn \cite{Tin1} or Cd \cite{Cd_iso} isotopes has provided the value for the asymmetry term of the nuclear incompressibility.
Although the theory has been well established to successfully reproduce the $K_\infty$ of $^{208}$Pb, it does not seem to work to treat the other open shell nuclei directly.
For example, theoretical calculations overestimate the peak energy in the Sn isotopes ($A =$ 112 - 124) above 1 MeV \cite{Tin1, Tin2} compared to the experimental data. 
This is the famous question: ``Why are Tins so soft?" \cite{PRC_Why}, which remains unanswered.
Nowadays, the development of new theories covering the open shell nucleus based on experimental data is one of the current directions in ISGMR research \cite{umesh}. 
Besides, the identification of ISGMR in neutron-rich nucleus, eventually weakly bound isotopes, is another hot topic \cite{umesh}, which can be accessed to extract the nuclear equation of state (EoS) of neutron matter at subsaturation density.

In this paper we have adjusted the effective interaction and introduced three different sets of Skyrme parameters, resulting in softer incompressibility coefficients. The verification of the potential parameters was focused on the ISGMR calculation.
The paper is structured as follows.
A brief review of the original EQMD model and the improvements are given in Sec. \ref{method}. 
The results and discussion are presented in Sec. \ref{result}, and a summary is given in Sec. \ref{conclusion}.

\section{Model and method \label{method}}

\subsection{The original EQMD model\label{EQMD}}

The EQMD model was developed to simulate low-energy nuclear reactions involving heavy systems with good computational performance.
Like most QMD-like transport approaches \cite{Aichelin,PRC_Deng,NST_Zhang,PRC_WangSS,NST_Zhang2,PRC_WangTT,NST_Xiao,Liu}, the nucleons in EQMD, i.e. protons and neutrons, are also assumed to be Gaussian wave packets. 
However, unlike most QMD-like models, the widths of the wave packets are propagated using the time-dependent variation principle (TDVP) \cite{FMD} rather than a constant as used in most QMD-like models. 
As mentioned in Ref.~ \cite{FMD}, this treatment in the trial state as a semiclassical, semiquantum mechanics involves more quantum effects than the fixed-width situation.
In the actual numerical calculation it makes the nuclear ground state much more stable \cite{papa}, which is advantageous for the description of giant resonance.

The wave function of the total system can be expressed as follows
\begin{equation}\label{wave}
\begin{aligned}
\Psi & =  \prod_i \varphi_i \left(\mathbf{r}_i\right) \\ & =\prod_i \left(\frac{v_i+v_i^*}{2 \pi}\right)^{3 / 4} \exp \left[-\frac{v_i}{2}\left(\mathbf{r}_i-\mathbf{R}_i\right)^2+\frac{i}{\hbar} \mathbf{P}_i \cdot \mathbf{r}_i\right],
\end{aligned}
\end{equation}
where $\mathbf{R}_i$ and $\mathbf{P}_i$ are the centers of the wave packet in coordinate and momentum spaces, and $v_i=\frac{1}{\lambda_i}+i\delta_i$ is the complex wave packet width corresponding  to the $i$th nucleon.
Following the TDVP, the propagation of phase space is expressed as four equations of motion as follows
\begin{equation}
\begin{aligned}
\dot{\mathbf{R}}_i & =\frac{\partial H}{\partial \mathbf{P}_i}+\mu_{\mathrm{R}} \frac{\partial H}{\partial \mathbf{R}_i}, ~~\dot{\mathbf{P}}_i=-\frac{\partial H}{\partial \mathbf{R}_i}+\mu_{\mathrm{P}} \frac{\partial H}{\partial \mathbf{P}_i}, \\
\frac{3 \hbar}{4} \dot{\lambda}_i & =-\frac{\partial H}{\partial \delta_i}+\mu_\lambda \frac{\partial H}{\partial \lambda_i}, ~~\frac{3 \hbar}{4} \dot{\delta}_i=\frac{\partial H}{\partial \lambda_i}+\mu_\delta \frac{\partial H}{\partial \delta_i}.
\end{aligned}
\end{equation}
Here $\mu_\mathbf{R}$, $\mu_\mathbf{P}$, $\mu_{\lambda}$ and $\mu_{\delta}$ are the friction coefficients. 
In the initialization phase, they are negative values to cool the system down to its (local) minimum point~\cite{EQMD} through energy dissipation. 
On the other hand, they remain strictly zero during the heavy ion reaction to satisfy the conservation of energy. 
Here $H$ is the Hamiltonian variable, expressed as
\begin{equation}
\begin{aligned}
H & =\left\langle\Psi\left|\sum_i-\frac{\hbar^2}{2 m} \nabla_i^2-\hat{T}_{\text {zero }}+\hat{H}_{\text {int }}\right| \Psi\right\rangle \\
& =\sum_i\left[\frac{\mathbf{P}_i^2}{2 m}+\frac{3 \hbar^2\left(1+\lambda_i^2 \delta_i^2\right)}{4 m \lambda_i}-\frac{t_i^\text{c.m.}}{M_i}\right]+H_{\mathrm{int}} .
\end{aligned}
\end{equation}
where the first three terms in brackets are the total kinetic energy of the $i$-th particle, $t_i^\text{c.m.}$ is the zero-point kinetic energy belonging to the $i$-th particle, which can be written as
\begin{equation}
t_i^\text{c.m.} 
 = \frac{\left\langle\phi_i\left|\hat{\mathbf{p}^2}\right| \phi_i\right\rangle}{2 m}-\frac{\left\langle\phi_i|\hat{\mathbf{p}}| \phi_i\right\rangle^2}{2 m}.
\end{equation}
$M_i$ is the so-called ``mass number" which can be expressed as follows: $M_i  =\sum_j F_{i j}$ \cite{EQMD} where 
\begin{equation}
\begin{aligned}
F_{i j} & =\left\{\begin{array}{cc}
1 & \left(\left|\mathbf{R}_i-\mathbf{R}_j\right|<a\right) \\
e^{-\left(\left|\mathbf{R}_i-\mathbf{R}_j\right|-a\right)^2 / b} & \left(\left|\mathbf{R}_i-\mathbf{R}_j\right| \geqslant a\right)
\end{array}\right.
\end{aligned}
\end{equation}
where the parameters $a$ = 1.7 fm and $b$ = 4 fm$^2$ in the original EQMD model.
The last $H_\text{int}$ is the total interaction energy for the whole system.
In the original version given by Maruyama {\it et al.} a very simple formula for the energy density function was proposed.
The mean field potential is written as 
\begin{equation}
H_{\text {int. }} = H_{\text{Sky. }}+H_{\text{Coul. }} + H_{\text{Sym. }} + H_{\text {Pauli }} .
\end{equation}
Here $H_{\text{Sky.}}$, $H_{\text{Coul.}}$, $H_{\text{Sym.}}$ and $H_{\text{Pauli}}$ represent the Skyrme, Coulomb, symmetry and Pauli interaction, respectively.
The forms of Skyrme interaction adopt the simplest one as follows
\begin{equation}
H_{\text{Sky. }}=\frac{\alpha}{2 \rho_0} \int \rho^2(\mathbf{r}) \mathrm{d} \mathbf{r}+\frac{\beta}{(\gamma+1) \rho_0^\gamma} \int \rho^{\gamma+1}(\mathbf{r}) \mathrm{d} \mathbf{r},
\end{equation}
where $\alpha$, $\beta$, $\gamma$ are the potential parameters which are listed in Table \ref{tab:table_EQMD}.
The symmetry term is written as 
\begin{equation}
H_{\text {Sym. }}=\frac{c_\text{s}}{2 \rho_0} \sum_{i, j \neq i} \int\left[2 \delta\left(T_i, T_j\right)-1\right] \rho_i(\mathbf{r}) \rho_j(\mathbf{r}) \mathrm{d} \mathbf{r} ,
\end{equation}
where $c_\text{s}$ is the coefficient of symmetry energy, $T_i$ and $T_j$ represent the isospin corresponding  to $i$th and $j$th nucleons, respectively.
The Pauli potential~\cite{EQMD} is written as 
\begin{equation} \label{eq:pauli}
H_\text{Pauli}  = \frac{c_P}{2} \sum_i\left(f_i-f_0\right)^\mu \theta\left(f_i-f_0\right), 
\end{equation}
where $c_p$ and $\mu$ are the strength and power of the Pauli potential, $\theta$ is the unit step function,  $f_i$ is the overlap of a nucleon with the same spin and isospin nucleons including itself, i.e. $f_i  \equiv \sum_j \delta\left(S_i, S_j\right) \delta\left(T_i, T_j\right)\left|\left\langle\phi_i \mid \phi_j\right\rangle\right|^2$, and $f_0$ is the threshold parameter, which takes a value close to 1.
When the $f_0=1$, the step function in Eq.~(\ref{eq:pauli}) can be ignored, since the $f_i-f_0$  always greater than 0.
The Pauli term can be understood as a repulsive force forbidding the nearby identical particle too close in the phase space. 
It makes the EQMD capability to describe $\alpha$ clustering structure in a nucleus.
Only the direct part of Coulomb interaction utilized in the EQMD written as 
\begin{equation}
H_\text{Coul.} = \frac{a_c}{2} \int \mathrm{d} \mathbf{r} \mathrm{d} \mathbf{r}^{\prime} \frac{\rho_{\mathrm{p}}(\mathbf{r})\rho_{\mathrm{p}}\left(\mathbf{r}^{\prime}\right)}{\left|\mathbf{r}-\mathbf{r}^{\prime}\right|}.
\end{equation}
Here $a_c = \frac{1}{137}$ is the fine structure constant.

\begin{table}[t]
\caption{\label{tab:table_EQMD}}
Parameters of QMD (soft), QMD (hard) and two sets of original EQMD model.
\begin{ruledtabular}
\begin{tabular}{lrrrr}
\textrm{}&
\textrm{QMD(soft)}&
\textrm{QMD(hard)}&
\textrm{EQMD 1}&
\textrm{EQMD 2}\\
\colrule
$\alpha$ (MeV)     & -356.0   & -124.0  & -116.6 &  -124.3\\
$\beta$ (MeV)      &  303.0   &  71.0   & 70.8   &    70.5\\
$\gamma$           &    7/6   &  2      & 2      &       2\\
$c_\text{s}$ (MeV) &      -   &  -      & 25.0   &    25.0\\
\end{tabular}
\end{ruledtabular}
\end{table}

In one of the earliest references to the QMD model~\cite{QMD_hard}, there are two sets of parameters with different values of incompressibility. 
The softer set, called QMD (soft), is associated with an incompressibility value of 200 MeV, and the harder one, called QMD (hard), is associated with a value of 380 MeV.
Table \ref{tab:table_EQMD} gives the parameter settings of the QMD (soft), the QMD (hard) and the EQMD model.
Obviously, the two sets of parameters proposed by Maruyama {\it et al.} are very different from the QMD (soft) parameters, but very close to the hard one, which is much larger than the current consensus range of $230 \pm 40$ MeV. 
Unfortunately, a soft incompressibility coefficient used in the EQMD model has been missing for a long time.
Another important aspect in the transport model is the binary collisions between nucleons. 
Some recent studies \cite{wangrui,xujun} have shown that the {\it NN} collision process seriously affects the magnitude of the damping of giant resonance (GR), but does not significantly affect the position of the peak energy.
In principle, both the mean-field nucleon propagation and the {\it NN} binary collisions should be considered simultaneously. 
In this article, we aim to resolve the discrepancy in the incompressibility coefficient between the EQMD model and the experimental data. 
For this reason, only the mean-field aspect is considered in this work to avoid further complications. 

\subsection{Modified mean-field potential \label{improve}}

Since the introduction of the Skyrme interaction, a large number of parameter sets have been developed that are consistent with the macroscopic constraints, such as the ground state properties of nuclei, the properties of nuclear matter, and so on.
It allows us to adopt the large number of Skyrme parameter sets already available, whose incompressibility values $K_\infty$ are well known.
Thus, the adoption of a standard Skyrme interaction can greatly simplify the adjustment of the potential parameters for our EQMD model.
Nowadays, a standard form of the Skyrme energy density functional is widely used by various QMD-like models, e.g. ImQMD~\cite{zhangyx}, LQMD~\cite{LQMD}, IQMD-BNU~\cite{IQMD_BNU} etc.

The effective interactions used in the EQMD model are quite simple according to current knowledge. 
Compared to the original EQMD model, we have adopted a more complete Skyrme type energy density potential functional \cite{skyrme} for our new version. 
It includes bulk, Coulomb, gradient terms and their symmetry part.
However, the momentum-dependent and spin-orbit terms are temporarily ignored due to their complicated forms. 
In this case the energy density functional is written as
\begin{widetext}
\begin{equation}
\begin{aligned}
U(\mathbf{r})&=\frac{1}{2} t_0\left[\left(1+\frac{1}{2} x_0\right) \rho^2-\left(x_0+\frac{1}{2}\right)\left(\rho_{\mathrm{n}}^2+\rho_{\mathrm{p}}^2\right)\right]
+\frac{1}{12} t_3 \rho^\alpha\left[\left(1+\frac{1}{2} x_3\right) \rho^2-\left(x_3+\frac{1}{2}\right)\left(\rho_{\mathrm{n}}^2+\rho_{\mathrm{p}}^2\right)\right] \\
+&\frac{1}{16}\left[3 t_1\left(1+\frac{1}{2} x_1\right)-t_2\left(1+\frac{1}{2} x_2\right)\right](\nabla \rho)^2
-\frac{1}{16}\left[3t_1\left(x_1+\frac{1}{2}\right)+t_2\left(x_2+\frac{1}{2}\right)\right]\left[\left(\nabla \rho_{\mathrm{n}}\right)^2+\left(\nabla \rho_{\mathrm{p}}\right)^2\right]\\
+&U_\text{Coul.}+U_\text{Pauli} .
\end{aligned}
\end{equation}
\end{widetext}
Here $t_0$, $t_3$ and $x_0$, $x_3$ are parameters of bulk-term, $t_1$, $t_2$ and $x_1$, $x_2$ are the parameters of gradient-term.
$U_\text{Coul.}$ and $U_\text{Pauli}$ are the energy density functional of Coulomb interaction and Pauli potential. 
The Coulomb term consisting of the direct and exchange terms \cite{coul_exchange} is written as 
\begin{equation}
U_\text{Coul.}(\mathbf{r}) = \frac{a_c}{2} \int \mathrm{d} \mathbf{r}^{\prime} \frac{\rho_{\mathrm{p}}(\mathbf{r})\rho_{\mathrm{p}}\left(\mathbf{r}^{\prime}\right)}{\left|\mathbf{r}-\mathbf{r}^{\prime}\right|}-\frac{3}{4} e^2\left(\frac{3}{\pi}\right)^{1 / 3} \rho_{\mathrm{p}}^{4 / 3}(\mathbf{r}).
\end{equation}
Here $a_c$ is the fine structure constant, the second term represents the exchange part. 
We keep the formulation of the Pauli potential as in Eq.~(\ref{eq:pauli}), but with an adjustment of the strength coefficient $c_\text{p}$. 

There are two ways to define the interaction parameters introduced in Ref.~\cite{EQMD}.
One is to retain the saturation state of nuclear matter with the Pauli potential, while adjusting the Skyrme interaction parameters to give 16 MeV binding energy at the saturation point. 
Another is to fix the original value of the Skyrme parameters with the adjustment of Pauli potential and zero point kinetic energy to satisfy the condition of the ground state of the nucleus, i.e., binding energy and nuclear charge radius.
We use the second method to define the interaction parameters in this article.
In addition, the effective mass of the Skyrme parameters should be as close as possible to 1.0 at saturation density, since the momentum-dependent and spin-orbit interactions are not taken into account here. 
These effects will be taken into account in the near future.
According to the above rules, three different parameter sets are given with $K_\infty$ from 200 to 268 MeV.
The detailed parameter settings are given in Table ~\ref{tab:table_skyrme}.
Some variables correlated with the symmetry energy, such as the slope ($L$), the curvature ($K_\text{Sym.}$) and the third derivative ($Q_\text{Sym.}$) are also listed.
In addition, the density dependent term is calculated exactly using the Monte Carlo integral. 
A similar method was first reported in \cite{three-body}.

\begin{table}[htbp]
\caption{\label{tab:table_skyrme}}
Skyrme parameters used in this work and the corresponding physics quantities. The $K_\text{Sym.}$ is the curvature of symmetry energy, and while  the $K_\infty$ is incompressibility.
\begin{ruledtabular}
\begin{tabular}{lrrr}
\textrm{}&
\textrm{SkP \cite{skp} }&
\textrm{SkT1 \cite{skt1}}&
\textrm{SkXce \cite{skxce}}\\
\colrule
$K_{\infty}$ (MeV)           & 200      &  236     & 268    \\
$m_0/m$                      & 1.00     &  1.00    & 0.99   \\
$L$ (MeV)                    & 19.6     &  56.18   & 33.48  \\
$K_\text{Sym.}$ (MeV)        & -266.6   &  -134.83 & -238.4 \\
$Q_\text{Sym.}$ (MeV)        & 508.4    &  319.0   & 356.9  \\         
$t_0$ (MeV fm$^5$)           & -2931.7  & -1794.0  & -1438.0\\
$t_1$ (MeV fm$^5$)           &  320.6   &  298.0   & 244.3  \\
$t_2$ (MeV fm$^5$)           & -337.4   & -298.0   & -133.7 \\
$t_3$ (MeV fm$^{3+3\alpha}$) &  18709.0 &  12812.0 & 12116.3\\
$x_0$                        &  0.292   &  0.154   & 0.288  \\
$x_1$                        &  0.653   & -0.500   & 0.611  \\
$x_2$                        & -0.537   & -0.500   & 0.145  \\
$x_3$                        &  0.181   &  0.089   & -0.056 \\
$\alpha$                     &  1/6     &  1/3     & 1/2    \\
$c_\text{p}$ (MeV)           &  18.6    &  16.8    & 17.6   \\
$f_0$                        &  1.0     &   1.0    & 1.0    \\
$\mu$                        &  1.3     &  1.3     & 1.3    \\
$a$ (fm)                     &  0.5     &  0.5     & 0.5    \\
$b$ (fm$^2$)                 &  4.0     &  4.0     & 4.0    \\
\end{tabular}
\end{ruledtabular}
\end{table}

\subsection{Giant monopole resonance \label{gmr}}

There are basically two groups of microscopic models that have been used for giant resonance calculations \cite{GiBUU}, i.e. a purely quantum mechanical approach and a semiclassical quantum microscopic transport model.  
The random phase approximation (RPA) method is a representative of the former approach \cite{NiuYF}, while quantum molecular dynamics (QMD) (-like), Boltzmann-Uehling-Ulenbeck (BUU) (-like), or Vlasov (-like) are the representatives of the other.
And they have been successfully applied to analyze the GR related topics. 
For example, the isospin-dependent quantum molecular dynamics model (IQMD) \cite{taochen} has been used to extract the isotopic dependence of the GMR peak energy, and the BUU has discussed the in-medium nucleon-nucleon ({\it NN}) cross section \cite{wangrui} as well as the effective mass effects in the GDR or GQR, etc. \cite{xujun}.

In a classical picture, GMR is understood as a radial collective vibration of an excited nucleus qualitatively, it is therefore called a breathing oscillation mode.
To excite a ground state nucleus to breathing oscillation, a perturbation can be added to the Hamiltonian of the nucleus at zero time, i.e., $\lambda\hat{Q}\delta(t-0)$. Here, $\hat{Q} = \frac{1}{A} \sum_i^{\mathrm{A}} \hat{r}_i^2$ is a suitable excitation operator \cite{wangr2} and $\lambda$ is a perturbation parameter.

In linear response theory, the response of excitation of $\Delta\hat{Q}$ is a function  of time which can be expressed as~\cite{wangr2}
\begin{equation}
\label{eq:std}
\begin{aligned}
\Delta\langle\hat{Q}\rangle(t) & = \left\langle f|\hat{Q}| f\right\rangle-\left\langle 0|\hat{Q}| 0\right\rangle \\
& =-\frac{2 \lambda \theta(t)}{\hbar} \sum_F|\langle F|\hat{Q}| 0\rangle|^2 \sin \frac{\left(E_F-E_0\right) t}{\hbar},
\end{aligned}
\end{equation}
where $|0>$ and $|f>$ are the nuclear states before and after perturbing at zero time, $|F>$ is the energy eigenstate of the excited nucleus with eigenenergy $E_F$, respectively. 
The strength function can be extracted as a Fourier integral of $\Delta\langle\hat{Q}(t)\rangle$ as
\begin{equation}
\label{eq:se}
S(E) = -\frac{1}{\pi \mu} \int_0^{\infty} \text{d} t \Delta\langle\hat{Q}(t)\rangle \sin \frac{E t}{\hbar},
\end{equation}
where $\mu$ is a scaling parameter.
Actually, when calculating the strength function with Eq.~(\ref{eq:se}), a damping factor \cite{wangrui, kong} of $e^{-\gamma t/\hbar}$ with $\gamma=1$ MeV is multiplied by the $\Delta\langle\hat{Q}\rangle(t)$, which is a common practice to avoid oscillations in the Fourier transformation due to the finite time span.
To overcome this drawback,
another method is to extract the periodic oscillation with the following form \cite{GiBUU,xujun}
\begin{equation}
\label{eq:sin}
Q(t) = -a  \sin \left( \frac{E_\gamma}{\hbar} t \right) \exp \left( -\frac{\Gamma }{2\hbar} t \right) + d,
\end{equation}
and the corresponding strength function can be integrated according to Eq.~(\ref{eq:se}).
Here $E_\gamma$ and $\Gamma$ are the GMR peak energy and width, respectively.

By analyzing the time evolution of the $\Delta\langle \hat{Q}\rangle(t)$ within the transport model, we can obtain the strength function and other quantities such as peak energy, width and energy weighted sum rules.
In general, the value of the scaling parameter $\mu$ in Eq.~(\ref{eq:se}) is equal to the value of the perturbed parameter.
This will affect the magnitude of the GMR strength.
However, in this work we focus on the energy peak, 
so the scaling parameter can be set to any value. 

\section{Results and Discussion \label{result}}

\begin{figure}[htbp]
\resizebox{8.6cm}{!}{\includegraphics{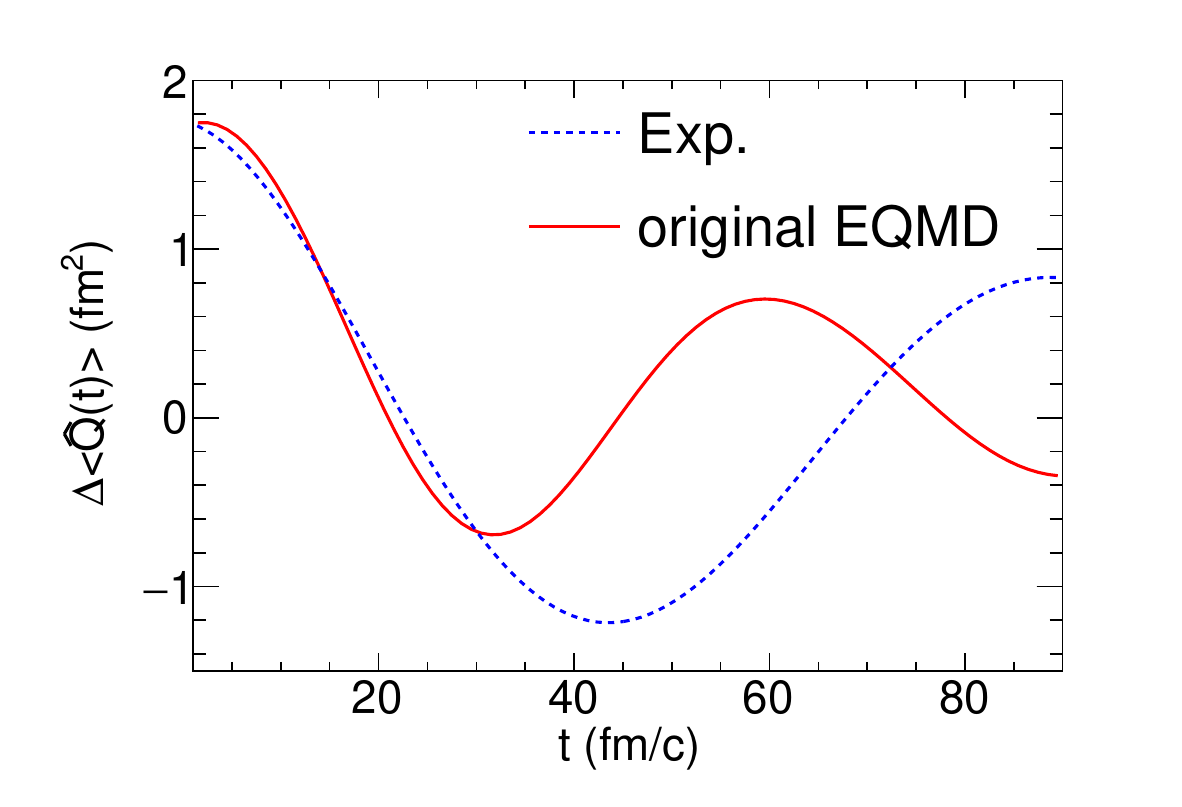}}
\caption{\label{fig:original EQMD} The $\Delta \langle \hat{Q} \left(t\right) \rangle$ plotted as a function of time. The red solid line is the result from the original EQMD model. And the blue line is derived from RCNP experiments.}
\end{figure}

To verify the newly embedded potential, we use the GMR to demonstrate its applicability.
To realize the giant monopole mode excitation at $t=0$ fm/$c$, the collective coordinate associated with the monopole vibration, a normalized scaled version of the nuclear density is given by the scaling relation \cite{scaling_relation}.
It is easy to obtain the new coordinates of the phase space after perturbing the $i$th nucleon as
\begin{equation}
\label{eq:perturbation}
\mathbf{R}_i \rightarrow c\mathbf{R}_i,~ \lambda_i \rightarrow c^2\lambda_i. 
\end{equation}
Here $c=1.03$ is a scaling parameter taken from Ref.~\cite{GiBUU}.
For this excitation the phase of $\langle\hat{Q}\rangle(t)$ is different from Eq.~(\ref{eq:sin}) with $\pi/2$. 
So the collective oscillation should be written as a cosine formation, i.e.,
\begin{equation}
\label{eq:cos}
Q(t) = a  \cos \left( \frac{E_\gamma}{\hbar} t \right) \exp \left( -\frac{\Gamma }{2\hbar} t \right) + d,
\end{equation}
and the corresponding strength function can be extracted as
\begin{equation}
\label{eq:se2}
S(E) = \frac{1}{\pi \eta} \int_0^{\infty} d t \Delta\langle\hat{Q}(t)\rangle \cos \frac{E t}{\hbar}.
\end{equation}

First, we plot the $\Delta \langle \hat{Q} \left( t \right) \rangle$ of the giant monopole for $^{208}$Pb as a function of time as calculated by the original EQMD model in Fig.~\ref{fig:original EQMD}.
The solid line is the result from the original EQMD model and the dashed line is derived from RCNP experiments \cite{patel} with $E_\gamma=13.7$ MeV and $\Gamma =3.3$ MeV according to Eq.~(\ref{eq:cos}). 
We artificially make these two curves equivalent at $t=0$ fm/$c$, which would not affect the position of the peak energy.  
In Fig. \ref{fig:original EQMD}, a period of oscillation of the $\Delta \langle \hat{Q} \left( t \right) \rangle$ derived from the experiments is plotted, which is about 90 fm/$c$.
In contrast, the frequency of the GMR oscillation in $^{208}$Pb calculated by the original EQMD model is about 1.5 times the experimental value, which means that $E_\gamma$ is about 20.55 MeV. 
This indicates that the incompressibility in the original EQMD model is indeed overestimated.

\begin{figure}[htbp]
\resizebox{8.6cm}{!}{\includegraphics{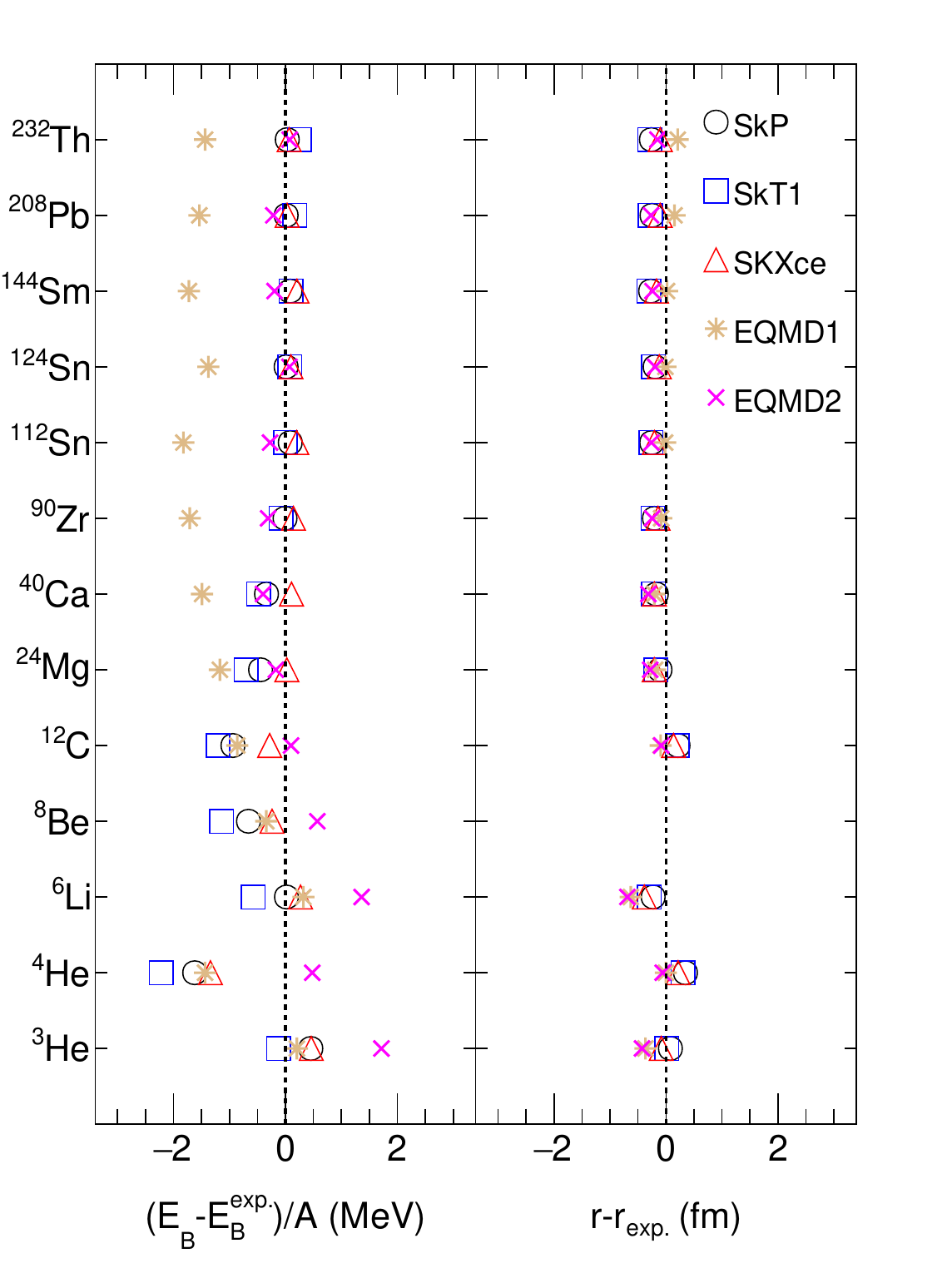}}
\caption{\label{fig:rms and ene vs mass num} The deviation of the calculated binding energies and radii of the nucleus from the experimental data as a function of mass number. Five sets of parameter are compared. }
\end{figure}

\begin{figure}[b]
\resizebox{8.6cm}{!}{\includegraphics{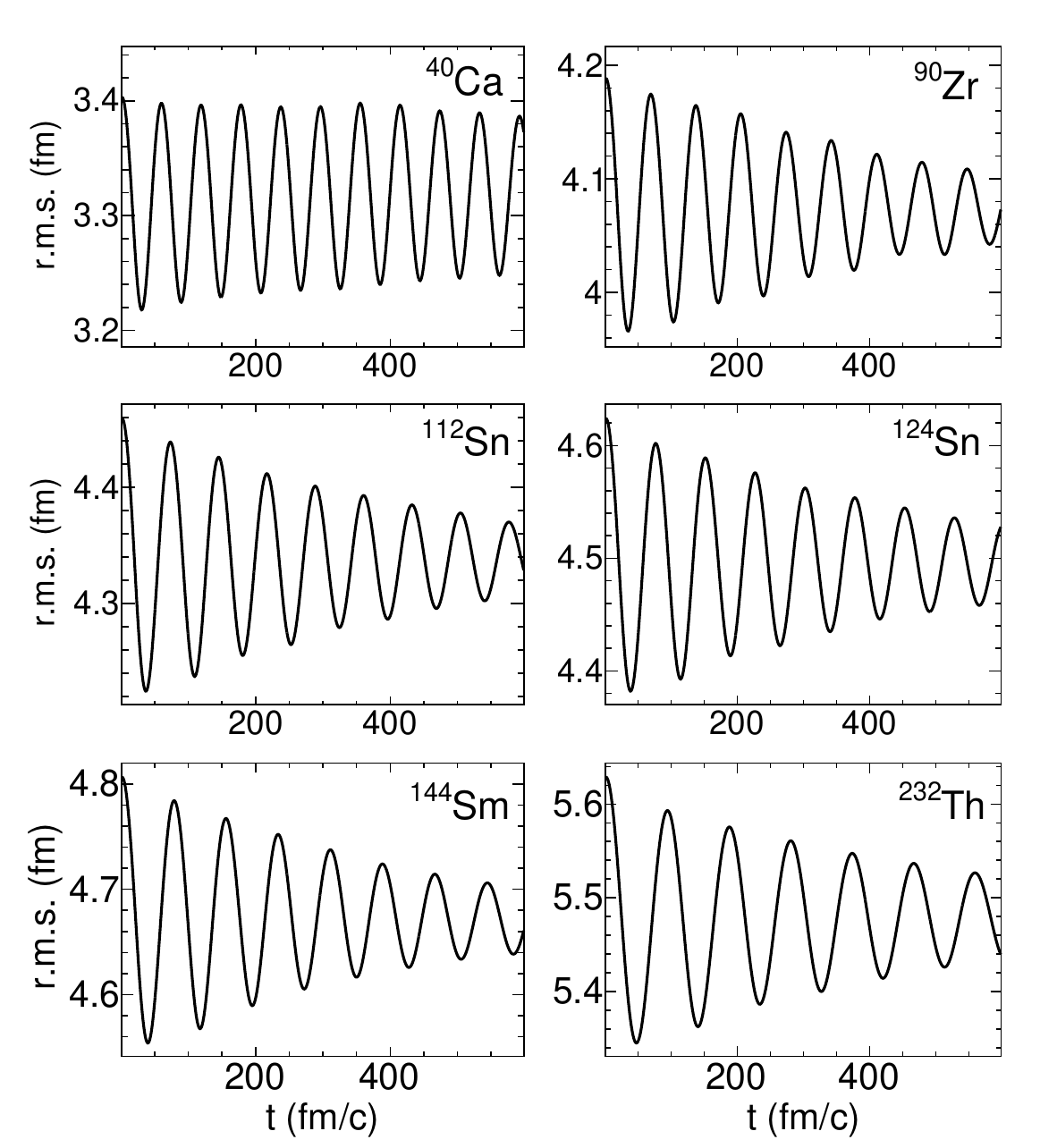}}
\caption{\label{fig:Ca to Th} The rms radii plot as a function of time for different nuclei with the mass numbers from 40 to 232.}
\end{figure}

Before starting to calculate the GMR oscillation, we also check the energy minimum states as initial ground states obtained by the frictional cooling method \cite{EQMD} in the framework of our EQMD model.
The deviation of the binding energies and radii of $^{3}$He, $^{4}$He, $^{6}$Li, $^{8}$Be, $^{12}$C, $^{24}$Mg, $^{40}$Ca, $^{90}$Zr, $^{144}$Sm, $^{208}$Pb and $^{232}$Th with five different parameter sets is shown in Fig. ~\ref{fig:rms and ene vs mass num}. 
The open circle, open block, and open triangle are the results simulated by the SkP, SkT1, and SkXce Skyrme parameters after frictional cooling, respectively. 
The detailed parameter settings and corresponding incompressibility values can be found in Table~\ref{tab:table_skyrme}.
For comparison, the same things derived from the original EQMD model with parameter sets 1 and 2 are plotted as stars and crosses in Fig.~\ref{fig:Ca to Th}.
As described in Maruyama's article~\cite{EQMD}, the original EQMD can reproduce the binding energy per nucleon reasonably well with parameter set 1 and almost perfectly with set 2 for most nuclei.
In our EQMD model, the deviations of binding energy per nucleon and radii are reasonably close to the experimental data for heavy ions with SkP, SkT1, and SKXce. 
However, the situation becomes complicated for nuclei lighter than $^{24}$Mg.
Especially for $^{4}$He the difference can reach about 2 MeV.
If one looks carefully, one can find a certain overestimation of the binding energies of $^3$He and $^{6}$Li in the case of EQMD parameter set 2. 
However, the deviations become very small in our EQMD model with SkP, SkT1 and SKXce.
On the other hand, with the exception of $^4$He, the binding energies for most nuclei can also be reasonably reproduced with the SKXce setting in this work.
Roughly speaking, based on the computational results, the SKXce and EQMD set 2 parameter settings can provide relatively reasonable reproductions of the nuclear ground state properties.

For the light nucleus, it shows a strong parameter dependence on the ground state reproduction.
In fact, this is not an uncommon problem for QMD-like models, and there have been some solutions to overcome this problem.
In the case of the ImQMD model~\cite{zhangyx}, the constant width of the wave packet is treated as a mass-number-dependent free parameter for fitting the ground state properties. 
However, this treatment cannot be adopted in our work because of the dynamic treatment of the wave packet width adopted in the EQMD model. 
In the case of the constrained molecular dynamics (CoMD) model~\cite{wangk_2}, the introduction of the Heisenberg principle can help to reasonably reproduce the binding energies and radii of light nuclei, which is a feasible solution for our EQMD model.
Despite this weakness for light nuclei, the correction for incompressibility within the EQMD model still represents a valuable step forward.
Based on the results, it may be more appropriate to focus initially on collisions involving heavy deformed nuclei, which has been a hot frontier in nuclear physics research recently.
The absence of momentum-dependent interactions prevents a large number of Skyrme-type parameter sets from being used in our model. 
We expect that the addition of momentum-dependent interactions and a large number of corresponding parameter sets will overcome this problem in the near future.

\begin{figure}[htbp]
\resizebox{8.6cm}{!}{\includegraphics{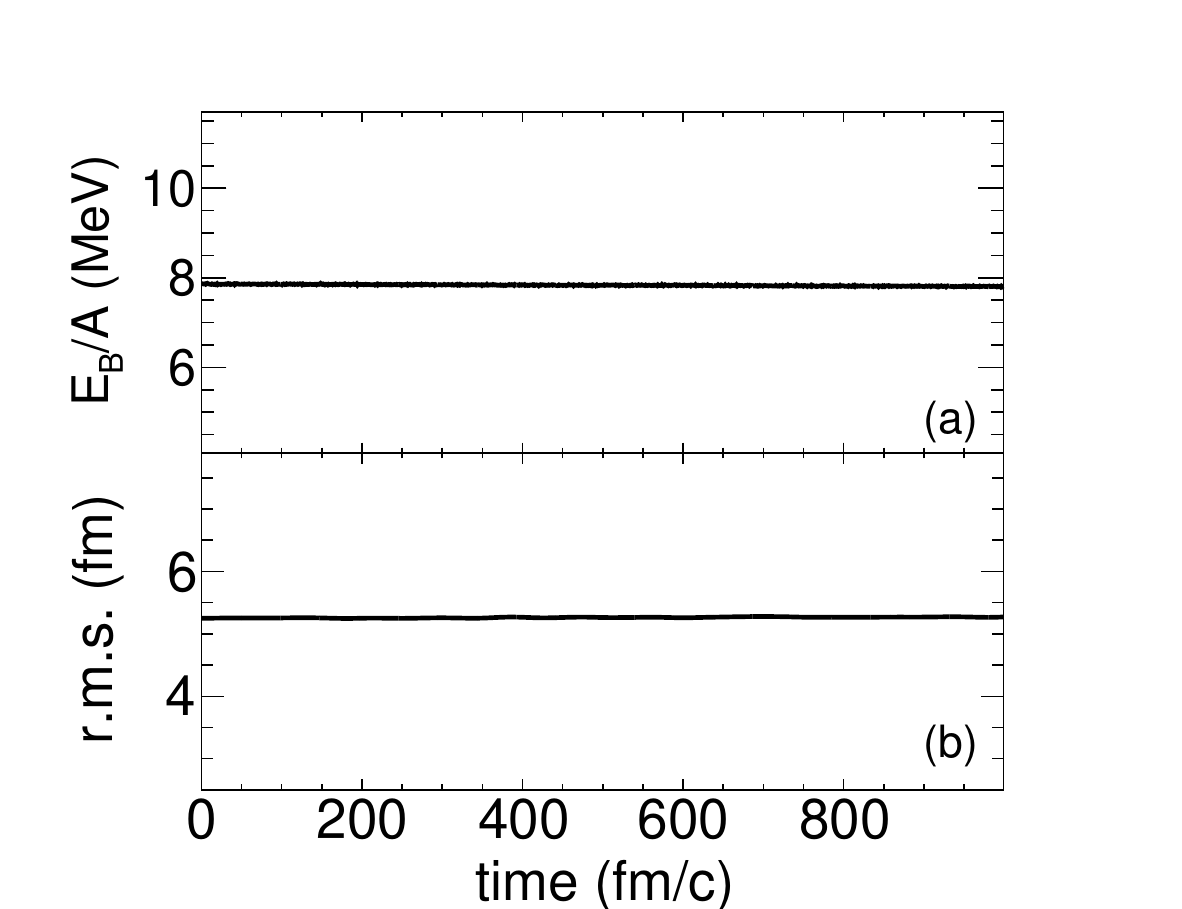}}
\caption{The time evolution of the binding energy (a) and r.m.s. charge radii (b) of the ground state $^{208}$Pb nucleus during 1000 fm/$c$.\label{fig:rms and ene vs time} }
\end{figure}

\begin{figure}[htbp]
\resizebox{8.6cm}{!}{\includegraphics{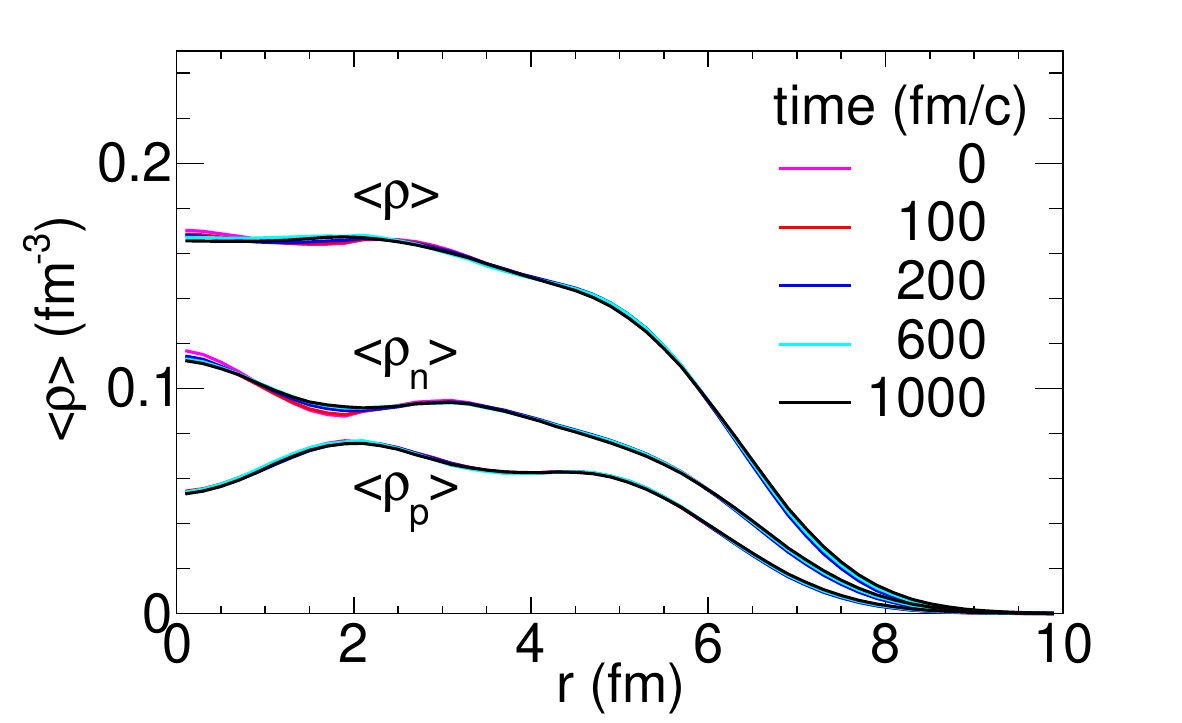}}
\caption{\label{fig:radii vs time}The time evolution of the radial density distribution of the ground state for $^{208}$Pb over 1000 fm/$c$ time span.}
\end{figure}

The stability of the ground state of the nucleus has a remarkable influence on the signal of the giant resonance \cite{GiBUU}, so it is necessary to initialize the neutron and proton densities according to a specific method, which can well improve the stability of the ground state \cite{GiBUU}.
For this reason, we also study the stability of the ground state evolution of the nucleus.
The time evolution of the binding energy and charge radii of $^{208}$Pb obtained by the SkP parameter setting during a thousand of fm/c span is shown in Fig.~\ref{fig:rms and ene vs time}. 
And the time evolution of the radial density distribution in $^{208}$Pb with the same parameter setting as in Fig.~\ref{fig:radii vs time}.
Both values of the binding energy, the charge radii and the radial density distribution remain almost stable over a long time, with only very small fluctuations that could be ignored.
The good stability of the ground states in the framework of our EQMD model is of great benefit for obtaining a clear signal of the ISGMR mode according to the previous research results \cite{GiBUU}.

\begin{figure}[t]
\resizebox{8.6cm}{!}{\includegraphics{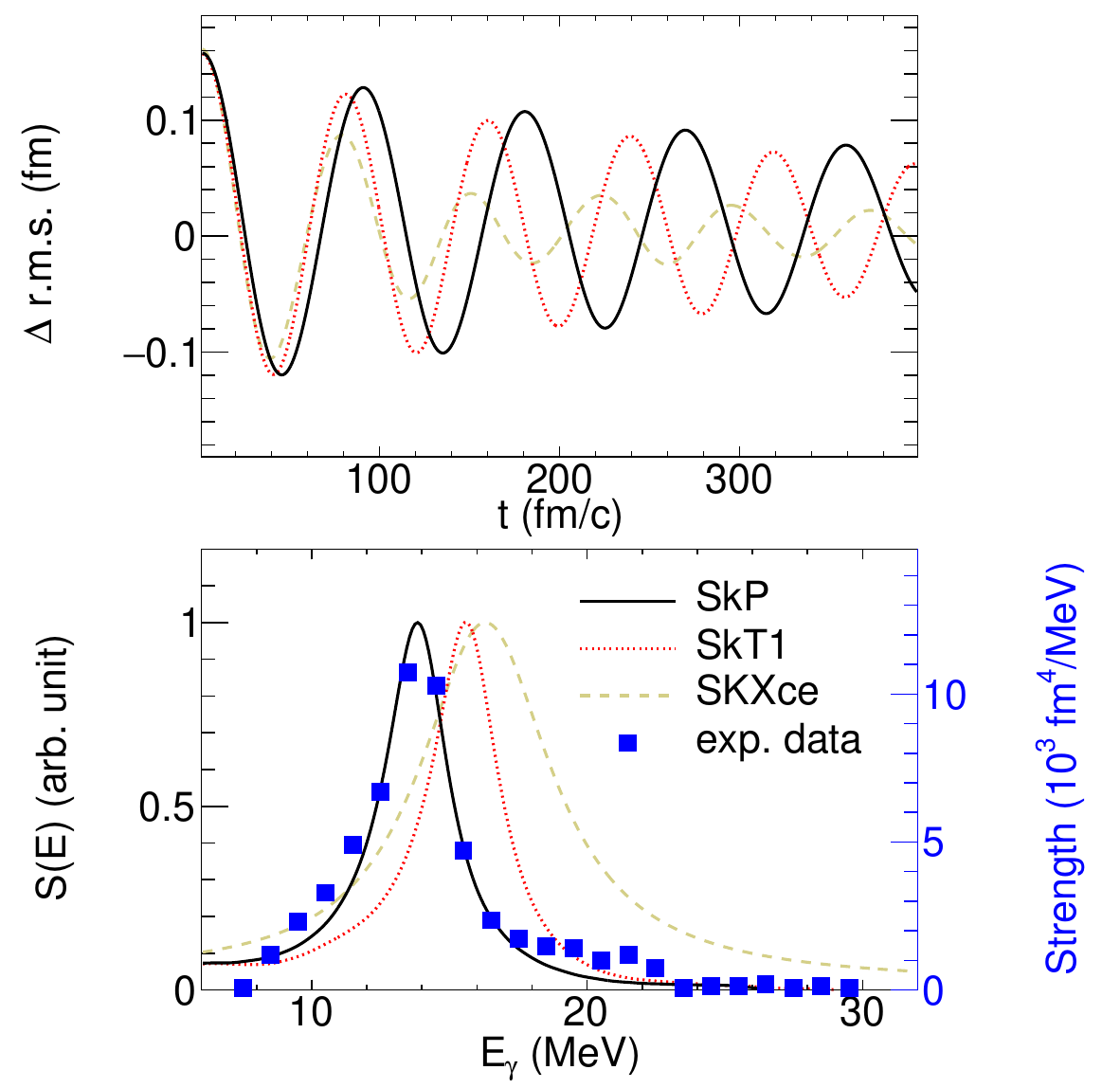}}
\caption{\label{fig:strength of pb}(a) Time evolution of the  $\Delta$rms(t)=rms(t)-rms(\text{g.s.}) for $^{208}$Pb using different sets of Skyrme parameters as indicated. (b) Excitation energy extracted from panel (a). The blue block represents the experimental data from RCNP \cite{umesh,patel}. } 
\end{figure}

The ISGMR in $^{208}$Pb has been used to investigate the value of the nuclear incompressibility in various theories and experiments because $^{208}$Pb is a double magic nucleus.
For this reason, we analyze the peak energy of the lead nucleus using the SkP, SkT1, and SkXce parameter sets, whose $K_\infty$ ranges from 200 to 268 MeV, to determine the appropriate setting for our model.  
In Fig.~\ref{fig:strength of pb}, the time evolution of the periodic oscillation at $\Delta$rms($t$)=rms($t$)-rms(gs) of $^{208}$Pb is plotted in panel (a), and the corresponding strength function extracted according to Eqs.~(\ref{eq:cos}) and (\ref{eq:se2}) is plotted in panel (b). Here, rms(g.s.) denotes the root mean square radius at the ground state. The solid line, dotted line, and dashed line represent the results using the SkP, SkT1, and SKXce parameter settings, respectively, and the blue block represents the experimental data in $^{208}$Pb for 386-MeV inelastic scattering data from the RCNP \cite{umesh,patel}.
According to Eq.~(\ref{eq:perturbation}), a same oscillation $c = 1.03$ on the radial density of $^{208}$Pb with different parameter setting is added at zero time, the periodic oscillation emerges three different frequencies. We extract the strength function from three different situations with a uniformly normalized amplitude, the peak energy position of three different setting sorts from small to large corresponding to the incompressibility from weak to strong as expected. In comparison with the experimental data, both the SkT1 and SKXce settings overestimate the peak energy for $^{208}$Pb to varying degrees, and the SkP parameter setting whose $K_{\infty}$ is close to 200 MeV shows the best agreement with the data.

\begin{figure}[htbp]
\resizebox{8.6cm}{!}{\includegraphics{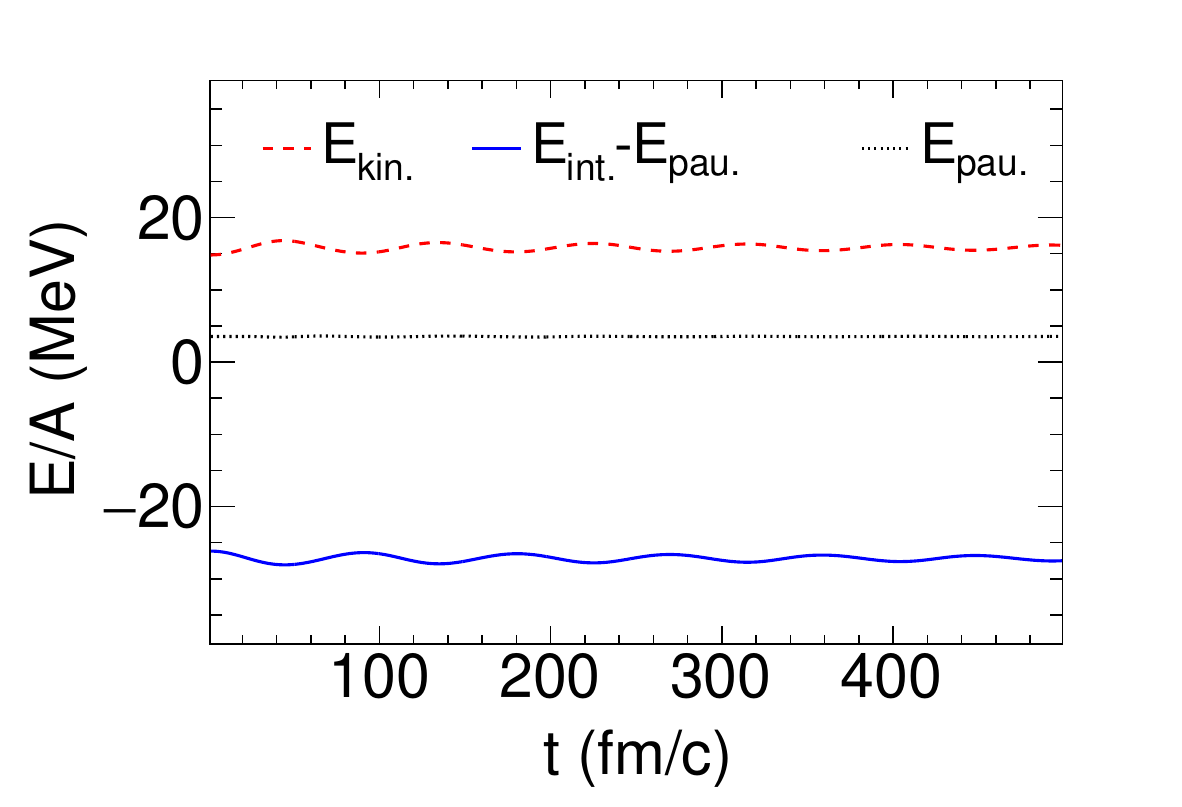}}
\caption{\label{fig:pauli}The kinetic energy, the total interaction energy excluding the Pauli potential, and the Pauli potential alone as a function of time with the SkP parameter setting for $^{208}$Pb.}
\end{figure} 

\begin{figure}[htbp]
\resizebox{8.6cm}{!}{\includegraphics{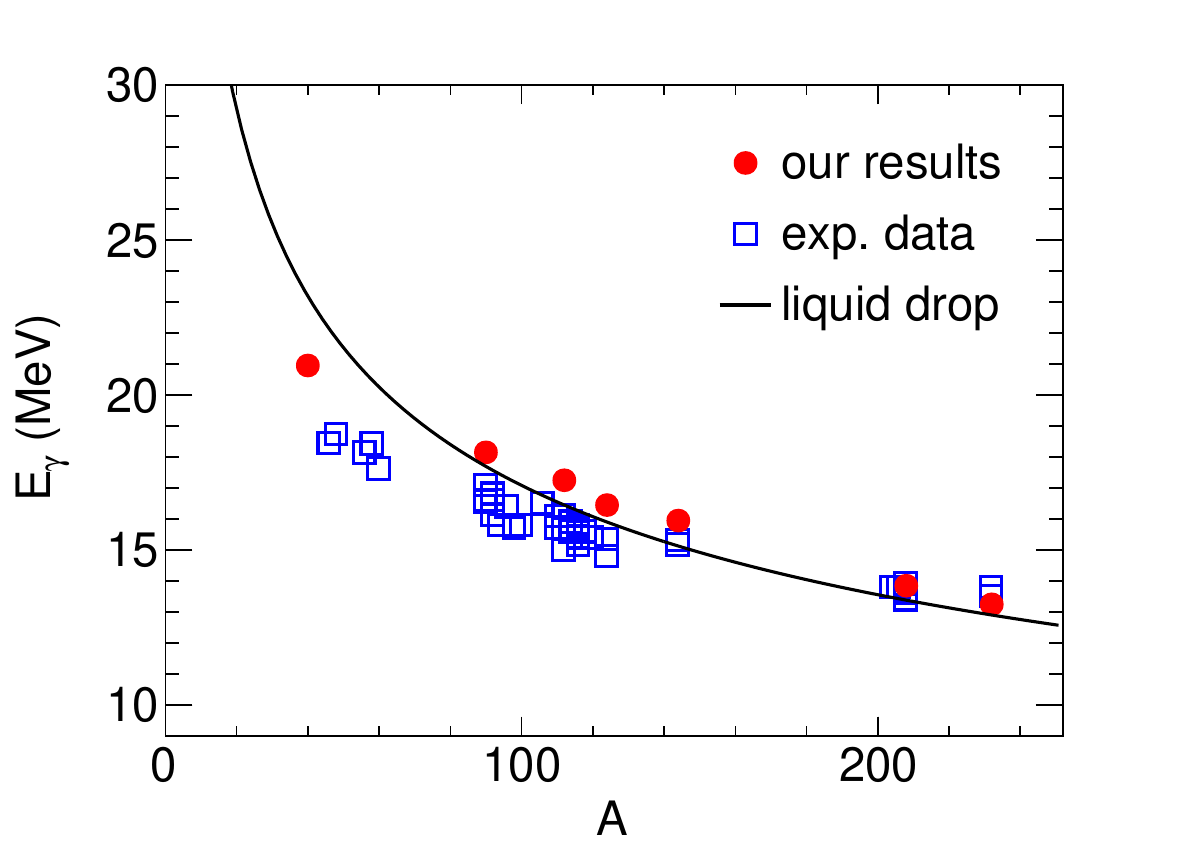}}
\caption{\label{fig:mass} Peak energies  of the GMR are  plotted  as a function of the mass number of nuclei. Our calculation results are marked as the full circles with the liquid drop formulation fit. The open blocks represent the experimental data from Ref.~\cite{patel}.}
\end{figure} 
The fermion properties of nucleons are very important for low energy nuclear reactions. 
However, only the AMD and FMD models treat the Pauli principle strictly with antisymmetrization of the phase space.
The method of strict antisymmetrization will seriously affect the computational efficiency.
So it takes a considerable amount of time to calculate the heavy nuclear system.
To mimic the effects of the Pauli principle with good performance, a phenomenological repulsive potential that prevents identical particles from coming close to each other in the phase space is adopted by Maruyama \cite{EQMD}, which is called Pauli potential.
In order to evaluate the effect of the Pauli potential during the GMR oscillation, we plot the kinetic energy, the total interaction energy excluding the Pauli potential, and the individual Pauli potential as a function of time for the GMR oscillation of $^{208}$Pb with the SkP setting in Fig.~\ref{fig:pauli}. 
Obviously, the absolute value of the Pauli potential is about 4 MeV/u, which is smaller than the total interaction energy, which is about 27 MeV/u. 
Moreover, the main energy exchange between the kinetic energy and the total interaction and the Pauli potential remains almost unchanged during the GMR oscillation.
In fact, the approach of the Pauli potential can not completely recover the Fermi properties for the nuclear matter \cite{FMD}.
How to deal strictly with the Fermi properties of nucleons while taking into account the efficiency is still a difficult task.

According to the above conclusions, the SkP parameter setting is reasonable to reproduce the breathing mode of $^{208}$Pb within the framework of our EQMD model.
We adopted this parametrization to study the mass dependence of the GMR peak energy.
The rms radii of $^{40}$Ca, $^{90}$Zr, $^{112}$Sn, $^{124}$Sn, $^{144}$Sm, and $^{232}$Th are shown as a function of time in Fig.~\ref{fig:Ca to Th}.
Again there is a clear oscillation signal over a long time.
The corresponding peak energy of the GMR as a function of mass number is plotted in Fig.~\ref{fig:mass}. 
The full circles are the results of our model, the open blocks are the experimental data \cite{patel}, and the solid line is the liquid-drop formulation calculation using $E_\gamma=\eta A^{-1 / 3}$ with $\eta=$79.3 MeV.
Here, $E_\gamma$ and $A$ are the peak energy and mass number, respectively.
To understand the dependence of the ISGMR peak energy ($E_\gamma$) on the mass, we use a formula like this \cite{KvsMass}
\begin{equation}
E_\gamma = \hbar \sqrt{\frac{K_A}{m\left\langle r^2\right\rangle}},
\end{equation}
where $m$ is the nucleon mass, $\langle r^2 \rangle$ is the mean square radius at the ground state which is proportional to $~A^{2/3}$, and $K_A$ is the finite nucleus incompressibility.
In the case of scaling model \cite{GiBUU}, it is roughly assumed that $K_A \approx K_\infty$, consequently the energy peak of the GMR follows the $A^{-1/3}$ law.
The present EQMD model can give a reasonable mass dependence of peak energy consistent with the experiment data in the region of heavy ion.
However, it overestimates about 1 MeV for the nuclei with mass number smaller than 100.
A weakness of our EQMD model is the lack of momentum dependent interaction due to its complex form, which currently prevents us from exploring the role of effective mass in this work.
In fact, in previous researches of our group \cite{taochen}, it was observed that the peak energy of GMR tends to increase when the momentum-dependent interaction is taken into account.
However, upon careful examination of the parameter settings, it is found that the bulk term parameters also undergo some changes, which also affect the GMR behavior, when the momentum-dependent interaction is included.
We believe that using the GMR to constrain the effective mass behavior will be a potential topic, and we are also considering adding this effect in the near future.

\section{Conclusion\label{conclusion}}

The EQMD model, as one of the few transport approaches, can be used to study cluster effects and heavily deformed nuclei in nuclear reactions with efficient computational performance.
However, for a long time only two sets of hard incompressibility parameters existed in EQMD. 
To address this issue, we introduce a standard Skyrme energy density function, such as the bulk, gradient term and the corresponding symmetry part, and the exchange term of the Coulomb interaction to replace the original mean-field form of the EQMD model in the present work.
In addition, we give three sets of Skyrme parameters, corresponding to the $K_\infty$ from 200 to 268 MeV, according to the ground state properties.
In the framework of our EQMD model, the SkP parameter setting can reproduce the GMR peak energy of $^{208}$Pb as the experimental data.
Based on the success of $^{208}$Pb, the GMR peak energies of $^{40}$Ca to $^{232}$Th have also been studied using the SkP parameter setting.
Overall, the results for the heavy nucleus are in good agreement with the experimental data, but overestimate the light nucleus by about 1 MeV.
The updated model can give a more reasonable EoS of nuclear matter, but the momentum-dependent interaction is still missing.
We will address this issue in our next work and study the effect of the momentum-dependent interaction for the giant resonance.

\section*{Acknowledgments}
This work is partially supported by the National Natural Science Foundation of China under Contracts No. $12347149$, $12147101$, $11890710$ and $11890714$,  the Strategic Priority Research Program of CAS under Grant No. XDB34000000, the National Key R\&D Program of China under Grant No. 2022YFA1602300, and the Guangdong Major Project of Basic and Applied Basic Research No. 2020B0301030008.

\end{CJK*}

\bibliography{eqmds_and_gmr}
\end{document}